\documentclass[twocolumn,prd,nofootinbib]{revtex4}
\usepackage{multirow}
\usepackage{amssymb}               
\usepackage{amsmath}
\usepackage{braket}
\usepackage{appendix}
\usepackage{hyperref}
\begin{document}

\title{Collapse of Coulomb Bound States of Vector Bosons}
\author{V. V. Flambaum$^{1}$}
\email{v.flambaum@unsw.edu.au}
\author{H. B. Tran Tan$^2$}
\affiliation{$^1$School of Physics, University of New South Wales,
Sydney 2052, Australia}
\affiliation{$^3$ Los Alamos National Laboratory, P.O. Box 1663, Los Alamos, New Mexico 87545, USA}

\begin{abstract}
  

Charged spin‑1 (vector) particles behave very differently from electrons or scalars in a Coulomb field. For an infinitely heavy \emph{point‑like} nucleus their bound‑state wavefunctions “fall to the centre,” and embedding the system in a renormalisable electroweak‑type theory does not remedy this short‑distance pathology. We therefore solve the pure Coulomb problem for a finite nuclear radius $R$ and recover the point‑nucleus limit by letting $R\to 0$. This approach allows us to include the crucial $\Upsilon$ term in the wave equations, which for the point-like nucleus is proportional to $\delta(r)$ and was ignored in the previous calculations of the energy spectrum. Several unusual effects emerge: (i) The $\Upsilon$ term supports a tower of states \emph{located mainly  inside the nucleus}. As $R \to 0 $ their number diverges, most lying in the negative‑energy continuum (emergy $\epsilon < - m c^2$). They trigger vacuum breakdown - particle‑antiparticle pair creation that ultimately screens the nuclear charge.
 (ii) Ordinary Sommerfeld-like states (with binding energy smaller $m c^2$) persist, but a finite fraction of each wavefunction leaks into the nucleus, even as $R \to 0$.
  (iii) Charge density of a \emph{negatively} charged vector particle  changes sign in a vicinity of the nucleus and becomes  \emph{positive} charge density, whereas the $\Upsilon$ term ensures its density inside the nucleus remains negative.
  (iv) For weak coupling,  $Z \alpha \ll 1$, yet with $mR <Z \alpha$, the “non‑relativistic’’ solution differs qualitatively from Schrödinger theory despite binding energies are well below $m c^2$; agreement is recovered only when $Z \alpha \ll  mR$.
  These phenomena highlight the distinctive and subtle behaviour of spin-1 particles in the Coulomb field.


\end{abstract}{}
\date{\today}
\maketitle

\maketitle

\section{Introduction}
The Proca-Maxwell equation \cite{Proca1936}, which describes the interaction between a massive spin 1 particle and the electromagnetic field, gives unacceptable results to the Coulomb problem, e.g., the scattering cross-section for large energy does not decrease with energy but tend to a constant~\cite{Corben1939,Oppenheimer1940,Tamm1940}. An attempt to amend this situation was made by Corben and Schwinger \cite{Schwinger1940}, who introduced to the interaction Lagrangian a new term which has the effect of forcing the particle's gyromagnetic factor to be $g=2$ (actually, $g=2$ is the gyromagnetic factor of $W$ boson in the Standard model). With this value of $g$, a physically acceptable spectrum was found to be given by the same Sommerfeld formula which describes the spectrum of the atomic electron, with appropriate values for the total angular momentum $j=0,1,...$ of a spin 1 particle.

The Corben-Schwinger formalism is not, however, free from flaws. In particular, it was found that for quantum states with $j=0$ and $j=1,l=0$, the charge of the particle near the Coulomb center becomes infinite, signifying the particle's fall to the center. Therefore, one could conclude  that the critical charge for the point-like nucleus corresponds to $Z\alpha=1$ for spin 1/2 particles, $Z\alpha=1/2$ for spin 0 particles and $Z\alpha=0$ for spin 1 particles. To resolve this issue, several authors have proposed further refinements \cite{pomeranskii1998,pomeranskii1999,pomeranskii2000} or completely new equations \cite{fushchich1985,fushchich1994} for the vector boson. 

Interesting features also include positive charge density of the negatively charged particle in a vicinity of the nucleus  \cite{Flambaum2007}. This  behavior, unique to a vector particle, is caused by its electric quadrupole moment and spin-orbit interaction, which give dominating contributions to the charge density  in the area of a rapid variation of the wave function. If the wave function is finite everywhere, the net effect of these  contributions to the integral of the charge density is zero. However, in the Coulomb case, the divergent integral of this singular wrong sign charge density makes the problem look like an apparent charge non-conservation.  

 Ref.~\cite{flambaum2006} examined the wave function collapse problem within a renormalisable electroweak‐type framework and showed that renormalisability alone does not ensure well-behaved solutions at short distances for non-perturbative problems such as bound states. In particular, the wavefunction of a charged vector particle collapses to a radius $r \sim 1/M$, where $M$ is the mass of the heavy Coulomb source - and $M$  can be arbitrarily large.

References \citep{Kuchiev2006,flambaum2006} demonstrated, however, that the Corben-Schwinger formalism  gives reasonable results, with the collapse problem disappearing if one takes into account the QED vacuum polarization from the beginning.  This is in stark contrast to the situation with scalar and spinor particles, where vacuum polarization acts only as a perturbation to the pure Coulomb field and thus has small effects on the particle's behavior. The difference lies in the fact that the equation governing a spin 1 particle contains a special term ($\Upsilon$ term) depending on the external current. With  QED vacuum polarization taken into account, this term produces a repulsive impenetrable potential near $r=0$ that prevents the particle from falling to the center. 
However, other electroweak vacuum polarization corrections, which appear in the Standard model (and any asymptotically free model), change the sign of the $\Upsilon$ potential from repulsive
to attractive  ~\cite{flambaum2006}. The collapse can be prevented by addition of light fermions, which switch the ultraviolet behavior of the theory from the asymptotic freedom to the Landau pole (like in pure QED).

  To investigate this behavior further, we follow another approach to the Coulomb problem, by considering the Coulomb center, henceforth called the nucleus, not as a point-like object but one with a finite size $R$.  Then we  recover the point-like nucleus limit of the problem  by taking $R\to 0$. In this finite $R$ case, the problem with infinite charge near the center does not persist, since at small distances, the pure Coulomb potential with its singularity at the origin is replaced with a regular potential produced by the spread nuclear charge distribution inside the nucleus. It is also of interest to ask what effects this `rounded-off' Coulomb potential has on the spectrum and wave function of the vector boson. Indeed, in atoms, the finite size of the nucleus is responsible for the familiar field isotope shift between energy levels of two atoms with nuclei of the same charge $Z$ but different mass numbers $A$.
    These nuclear-size corrections to the energy levels and wave functions of the electron may be calculated accurately with the use of normal perturbation theory,  at least in the non-relativistic limit 
  (the relativistic case indicates some deviation from the simple perturbation theory approach due to the singularity of the Dirac wave function at the origin in the point-like nucleus case, see e.g. \cite{flambaum2018}). In contrast, the finite nuclear radius, even an extremely small one, produces radical changes in the results for a spin 1  particle.


 To avoid misunderstanding, we do not consider here any electroweak and finite nuclear mass effects. We are solving  a purely theoretical problem of a light  charged vector particle in the field of the Coulomb center of finite size, focusing on the most interesting case  $mR \ll  Z\alpha \ll 1$. Note that the Coulomb  potential inside the nucleus in this case exceeds mass, $Z\alpha/R > m$. This is not a problem for spin 1/2 and spin 0 particles, the Schrodinger equation still gives correct results. However, for spin 1 particles a "non-relativitic" solution for small interaction constant $Z \alpha$,  for $mR < Z\alpha \ll 1$  differs from the solution obtained using  the Schrodinger equation, in spite of the binding energy $E_b \ll mc^2$ for the Schrodinger-like energy levels (there is agreement with the Schrodinger equation results in the true non-relativistic case $Z\alpha/R \ll  m$).  

 The finite nucleus problem may also be relevant to behavior of a vector particle in the field of a small charged black hole.
 

This paper is organized as follows. In Sec.~\ref{Theory}, for the purpose of fixing notations, we represent the Corben-Schwinger equation for a vector boson interacting with a static electric field. In Sec.~\ref{Correction to wave function and energy}, we find the energy  of the vector boson  for the finite size of the nucleus, using  Wentzel–Kramers–Brillouin (WKB)  semiclassical method. We also discuss peculiar behavior of the charge density and the effects of the vacuum polarization potentials.
A summary of the findings is presented in Sec.~\ref{Conclusion}.  For comparison, in the Appendix we presented the magnetic porlarization case where the wave function does not collapse and a finite nuclear radius $R$ yields only small corrections; the wave equation in this case coincides with the Klein-Gordon equation for spin zero paritcles. Throughout the paper, the natural units $\hbar=c=1$ and $e^2=4\pi\alpha$ are used, where $e<0$.

\section{Wave equations for vector partcle}\label{Theory}

The Lagrangian for a vector boson of mass $m$ and charge $e$ propagating in an external electromagnetic field has the form \cite{Corben1939}
\begin{equation}\label{Lagrangian}
\begin{aligned}
\mathcal{L}&=-\frac12\left(\nabla_{\mu}W_{\nu}-\nabla_{\nu}W_{\mu}\right)^\dagger\left(\nabla^{\mu}W^{\nu}-\nabla^{\nu}W^{\mu}\right)\\
&+ieF^{\mu\nu}W_{\mu}^\dagger W_{\nu}+m^2W_{\mu}^\dagger W^{\mu}\,,
\end{aligned}
\end{equation}
where $W_{\mu}$ is the vector boson field, $F_{\mu\nu}\equiv\partial_{\mu}A_{\nu}-\partial_{\nu}A_{\mu}$ is the electromagnetic field tensor and $\Delta_{\mu}\equiv\partial_{\mu}+ieA_{\mu}$ is the covariant derivative. Here, $A_{\mu}$ is the electromagnetic 4-potential.

The Euler-Lagrange equation resulting from the Langrangian~\eqref{Lagrangian} reads
\begin{equation}
\label{EulerLagrange}
(\nabla^2+m^2)W^{\mu}+2ieF^{\mu\nu}W_{\nu}-\nabla^{\mu}\nabla^{\nu}W_{\nu}=0\,,
\end{equation}
which, after taking covariant derivative, gives
\begin{equation}\label{Constraint}
m^2\nabla_{\mu}W^{\mu}+iej_{\mu}W^{\mu}=0\,,
\end{equation}
where $j^{\mu}\equiv\partial_{\nu}F^{\mu\nu}$ is the external current that produces the potential $A_{\mu}$.

In the static case where $j^{\mu}=(\rho({\bf r}),\bf{0})$, one can assume that 
\begin{equation}
\nabla_0^2W_{\mu}=-(\epsilon-U)^2W_{\mu}\,,
\end{equation}
where $\epsilon$ is the vector boson's energy and $U({\bf r})$ is its potential energy. Equation \eqref{Constraint} then becomes
\begin{equation}\label{Constraint2}
iW_0=(\epsilon-U-\Upsilon)^{-1}\nabla\cdot{\bf W}\,,
\end{equation}
where $W_0$ and $\bf {W}$ is the temporal and spatial components of $W^{\mu}$ and $\Upsilon$ is defined by
\begin{equation}\label{UpsilonDef}
\Upsilon\equiv e\rho/m^2=-\Delta U/m^2\,.
\end{equation}

Inserting Eq.~\eqref{Constraint2} into Eq.~\eqref{EulerLagrange}, one finds the equation
\begin{equation}\label{FinalLagrange}
\left[(\epsilon-U)^2-m^2\right]{\bf W}=-\Delta{\bf W}-2w\Delta U-\nabla(\Upsilon w)\,,
\end{equation}
where $w\equiv iW_0$.
For later use, we note that the charge density of the vector boson is given by
\begin{equation}\label{RhoDef}
\begin{aligned}
\rho^W&=2e\left[(\epsilon-U)({\bf W}^\dagger\cdot{\bf W}+w^\dagger w)\right.\\
&\left.+{\bf W}^\dagger\cdot\nabla w+{\bf W}\cdot\nabla w^\dagger-\Upsilon w^\dagger w\right]\,.
\end{aligned}
\end{equation}

If we assume that the potential $U$ is spherical then the total angular momentum $j$ of the vector particle is conserved and its (vector) wave function $\bf W$ may be separated into magnetic, electric and longitudinal components. Also, due to their different parities, the magnetic polarization decouples from the electric and longitudinal polarizations. As a result of this decoupling, we may consider independently the magnetically polarized mode where
\begin{equation}\label{MagDef}
{\bf W}=f(r){\bf Y}^{(m)}_{jm}\,,
\end{equation}
and the electrolongitudinally polarized mode where
\begin{equation}\label{ElectroLongDef}
{\bf W}=u(r){\bf Y}^{(e)}_{jm}+v(r){\bf Y}^{(l)}_{jm}\,.
\end{equation}
Here, $f(r)$, $u(r)$, and $v(r)$ are radial wave functions and ${\bf Y}^{(m,e,l)}_{jm}$ are vector spherical harmonics, defined as
\begin{subequations}
\begin{align}
{\bf Y}^{(e)}_{jm} & \equiv \nabla_{\bf n}Y_{jm}({\bf n})/\sqrt{j(j+1)}\,,\\
{\bf Y}^{(m)}_{jm} & \equiv {\bf n} \times {\bf Y}^{(e)}_{jm}\,,\\
{\bf Y}^{(l)}_{jm} & \equiv {\bf n} Y_{jm}({\bf n})\,,
\end{align}
\end{subequations}
with ${\bf n}={\bf r}/r$ being the unit vector in the direction of $\bf r$, $Y_{jm}$ the spherical harmonics, and $\nabla_{\bf n}$ the angular part of the gradient operator, i.e., $\nabla F({\bf n})=\nabla_{\bf n}F({\bf n})/r$. The magnetic and electrolongitudinal modes, corresponding to the terms ${\bf Y}^{(l)}_{jm}$ and ${\bf Y}^{(e)}_{jm}$ in Eqs.~\eqref{MagDef} and~\eqref{ElectroLongDef}, are only defined for $j>0$. For $j=0$, only the purely longitudinal mode, corresponding to the term $Y_{jm}({\bf n})$ in Eq.~\eqref{ElectroLongDef}, exists.

Below, we will solve Eq.~\eqref{FinalLagrange} in the finite nucleus  potential for the  purely longitudinal polarizations ($j$=0), which is the most interesting case. 


\section{Longitudinal polarization,  \texorpdfstring{$j=0$}{} } \label{Correction to wave function and energy}
\subsection{Wave equation for longitudinal polarization}
In the case where $j=0$, the wave function $\bf W$ is purely longitudinal and we have
\begin{equation}\label{LongDef}
{\bf W}=v(r){\bf n}\,,
\end{equation}
which, when substituted into Eq.~\eqref{FinalLagrange}, gives the following equation for $v$
\begin{equation}\label{LongEqn}
\frac{d^2v}{dr^2} + G\frac{dv}{dr} + H v=0\,,
\end{equation}
where
\begin{equation}\label{GH}
\begin{aligned}
G=\frac{2}{r}+\frac{U'}{\epsilon-U}+\frac{U'+\Upsilon'}{\epsilon-U-\Upsilon},\\
H=\frac{(\epsilon-U-\Upsilon)\left[(\epsilon-U)^2-m^2\right]}{\epsilon-U}\\
-\frac{2}{r}\left(\frac1r-\frac{U'}{\epsilon-U}-\frac{U'+\Upsilon'}{\epsilon-U-\Upsilon}\right).
\end{aligned}
\end{equation}
%
For the solution outside the nucleus, where $U=-\zeta/r$ ($\zeta= Z \alpha$) and $\Upsilon=0$, Eq.~\eqref{LongEqn} reduces to
\begin{equation}\label{LongEqnOut}
\begin{aligned}
&\frac{d^2v}{dr^2}+2\left(\frac{1}{r}+\frac{U'}{\epsilon+\zeta/r}\right)\frac{dv}{dr}\\
&+\left[\left(\epsilon+\frac{\zeta}{r}\right)^2-m^2-\frac{2}{r^2}+\frac{4U'}{r(\epsilon+\zeta/r)}\right]v=0\,.
\end{aligned}
\end{equation}
We follow Ref.~\cite{Kuchiev2006} and introduce the dimensionless variable $x\equiv \epsilon r/\zeta$ and perform the transformation
\begin{equation}
v=\frac{1+x}{x^2}\left(\frac{d}{dx}+L\frac{1+x}{x}-\frac{1}{1+x}\right)\tilde{\varphi}\,,
\end{equation}
with which Eq.~\eqref{LongEqn} becomes
\begin{equation}\label{LongReducedEqn}
\frac{d^2\tilde{\varphi}}{dx^2}+\left[\frac{2\zeta^2}{x}-\frac{L(L+1)}{x^2}\right]\tilde{\varphi}=\eta^2\tilde{\varphi}\,,
\end{equation}
where $\eta\equiv \zeta\sqrt{m^2-\epsilon^2}/\epsilon$ and $L\equiv\gamma+1/2=\sqrt{1/4-\zeta^2}+1/2$. Equation \eqref{LongReducedEqn} is a Whittaker equation which has solution
\begin{equation}
\tilde{\varphi}=c_1M_{\frac{\zeta^2}{\eta},L+\frac12}(2\eta x)+c_2W_{\frac{\zeta^2}{\eta},L+\frac12}(2\eta x)\,.
\end{equation}

In the case of a point-like nucleus, the solution $W_{\frac{\zeta^2}{\eta},L+\frac12}(2\eta x)$ is singular at $r=0$ and must thus be excluded. The requirement that $\tilde{\varphi}$ be regular at infinity forces the eigenvalue $\eta$ to take the form
\begin{equation}\label{LongSommerfeld}
\eta=\frac{\zeta^2}{L+n-1}\,,\quad\quad n=2,3,...\,,
\end{equation}
which gives the energy spectrum \cite{Schwinger1940}
\begin{equation}\label{eq:Sommerfeld_j=0}
\epsilon=\epsilon_{n0}=m\left[1+\frac{\zeta^2}{(\gamma+n-1/2)^2}\right]^{-1/2}\,.
\end{equation}
This result looks similar to the Sommerfeld formula for fermions. The difference is that in our case $j=0$. 

 Note that this solution does not include contribution of $\Upsilon$, which for the point-like nucleus is proportional to $\delta(r)$ ( the contact interaction).
 As we demonstrate below, incorporating the $\Upsilon$ contribution profoundly alters the result.

\subsection{Semiclassical (WKB) solution for the finite nucleus case}
We consider solution for the longitudinal  mode $j=0$ which incudes effects of $\Upsilon$ potential localized inside the nucleus. We can eliminate first derivative of $v$  using substitution \cite{Kuchiev2006} 
\begin{equation}\label{vphi}
v=\frac{1}{mr}[(\epsilon-U)(\epsilon-U -\Upsilon)]^{1/2}\phi\,.
\end{equation}
 Then equation for $\phi$ gets a Schrodinger-like form
\begin{equation}\label{eqphi}
-\frac{d^2 \phi}{dr^2} + V \phi=0\,,
\end{equation}
where 
\begin{equation}\label{V}
 V=-H +\frac{1}{4}G^2 + \frac{1}{2} \frac{dG}{dr}\,
\end{equation}
Note that in the non-relativistic case ($E=\epsilon -m$, $m \gg (E -U)$, $m \gg \Upsilon $),  Eq. (\ref{vphi})  is reduced to $v=\phi/r$, which is a conventional substitution transforming a radial equation into a Schrodinger one-dimensional equation with the effective potential $U_{eff}= U + l(l+1)/(2 mr^2) $.

Semiclassical WKB solution of a one-dimensional Schrodinger equation is presented as 
\begin{equation} \label{vclassical}
\phi \sim \frac{1}{ \sqrt{p}}\sin{(\int_a^{r} p dr + \theta_0)},
\end{equation}
where $\theta_0 \lesssim 1$, $p$ is the classical momentum and $a$ is the classical turning point, where $p=0$.

 Let us first consider the non-relativistic Coulomb energy levels $E_n=\epsilon_n -m $, basing on the Bohr-Sommerfeld quantization rule which follows from the boundary conditions for $\phi$ in Eq.(\ref{vclassical}) : 
\begin{equation} \label{Bohr}
\int_a^b p dr= \pi n +\theta,
\end{equation}
where $\theta \lesssim 1$, and $a,b$ are  the classical turning points, where $p=0$; on the distance $r>b$ the wave function decays exponentially.

For an illustration, we firstly  find from this equation the ordinary  non-relativistic Coulomb energy levels  for the point-like nucleus case. We should assume the non-relativistic energy $E \ll m $ and the potential $U \ll m$. This gives us conventional Schrodinger equation for $\phi$. Semiclassical approximation assumes that momentum  is large ($pr \gg 1$) and its derivative is relatively small. Keeping only the dominating term $\frac{d^2 \phi}{d r^2} \propto p^2$ in Eq.  \ref{eqphi}, we obtain the conventional expression for the classical momentum $p= \sqrt{2 m (E-U_{eff})}$, where $U_{eff}= - Z \alpha/r + l(l+1)/(2 mr^2) $. The orbital angular momentum is  $l=1$ for $j=0$ and spin 1.

The integration in Eq. (\ref{Bohr}) gives  
\begin{equation} \label{BohrCoulomb}
\pi \left[Z\alpha \sqrt{\frac{m}{2 (-E_n)}} - \sqrt{l(l+1)}\,\right]\approx  \pi n_0,
\end{equation}
This gives  Coulomb energy 
\begin{equation} \label{CoulombEnergy}
E_n=- \frac{m Z^2 \alpha^2}{2 n^2}\,,
\end{equation}
where the principal quantum number $n \approx n_0 + \sqrt{l(l+1)}$. There is a small difference with accurate expression  $n=n_0 +l$.  However, the Bohr -Sommerfeld  rule  gives  accurate results for $n \gg 1$, where the  WKB approximation is formally applicable. Moreover, for $n \gg l$ we may simply neglect the centrifugal  term contribution to the energy. 

Now consider contribution of the finite nuclear size to the Bohr-Sommerfeld integral Eq. (\ref{Bohr}):  
\begin{equation} \label{BohrIn}
\int_0^{R} (p_i - p_c) dr,
\end{equation}
where $p_i$ is the classical momentum within the nucleus and $p_c =\sqrt{2 m (E + Z\alpha/r}$ is the classical momentum in the Coulomb field of the point-like nucleus, which we should subtract since we have already included its contribution to Eq. (\ref{BohrCoulomb}). 

We will consider the most interesting case $mR \ll  Z \alpha \ll 1$. This condition gives us $m \ll U \ll \Upsilon$ inside the nucleus. The case of the nonrelativistic Coulomb energies $E_n \ll m \ll U \ll \Upsilon$, which we are considering in this subsection, gives us an additional energy-independent contribution 
\begin{equation} \label{BohrDelta}
\int_0^{R} (p_i - p_c) dr \equiv \pi \Delta ,
\end{equation}
to the Bohr-Sommerfeld  equation 
\begin{equation} \label{BohrCoulombDelta}
\pi \left(\Delta +Z\alpha \sqrt{\frac{m}{2(- E_n)}}\right )= \pi n,
\end{equation}
This gives a corrected energy 
\begin{equation} \label{CoulombEnergy}
E_n=- \frac{m Z^2 \alpha^2}{2 \nu^2}.
\end{equation}
where $\nu=n-\Delta$ is the effective principal quantum number. We may say that the levels are re-numbered. The magnitude of $\Delta =n- \nu $ indicates the number of orbitals with the same angular quantum numbers ($j=0$ for the longitudinal polarization) inside the nucleus, $n_{in} \sim \Delta$.

Parameter $\Delta$ and the effective principal quantum number $\nu=n-\Delta$  are  not expected to be integer numbers, so the energy levels do not coincide with the energy levels in the Coulomb potential with a point-like nucleus. However, the centrifugal barrier for $l=1$ separates two areas of the classical radial motion,  outside the nucleus and inside the nucleus, so there are two nearly independent contributions to the phase.   This may reduce the difference with the pure Coulomb spectrum after the renumbering of the levels. 

This formula (\ref{CoulombEnergy}) is similar to the Rydberg formula for the energy of the external electron traveling far from atomic core in the Coulomb field of the remaining ion. The explanation is similar. The energy of the external electron $E_n$ is negligible within the electron core, $E_n \ll U$. Therefore, the phase shift $\pi \Delta$ is energy-independent. Taking into account that the  potential for external electron outside the core is $U= -(Z-N+1) \alpha/r$, we obtain  Rydberg formula for external  electron  energy 
\begin{equation} \label{RydbergEnergy}
E_n^{Rydberg} =- \frac{m (Z-N +1)^2 \alpha^2}{2 \nu^2},
\end{equation}
where $N$ is the total number of electrons (for a neutral atom $N=Z$), and $\nu=n-\Delta$ is the effective principal quantum number. The magnitude of $\Delta =n- \nu $ indicates the number of orbitals with the same angular quantum numbers inside the electron core, $n_{core} \sim \Delta  -n_{min}$,
where $n_{min}=l+1$ is the minimal principal quantum number for a given orbital angular momentum $l$.  
Typical value of $\nu$ for the lowest external electron level in a neutral atom is given by   $1< \nu <2$. 

We may say, that we, in fact,  shifted boundary condition for the  external electron from $r=0$ to the boundary of the electron core $R_c \sim a_B$ ($a_B$ is the Bohr radius) and then solve the ordinary Coulomb problem. 
In our spin 1 particle case, we may provide a similar qualitative explanation for the Rydberg-type  equation (\ref{CoulombEnergy}) for energy levels: we effectively shifted boundary condition from $r=0$ to the nuclear surface $r=R$ and then solve  the Coulomb problem in the potential $U = -  Z\alpha/r$ with this new boundary condition at $r=R$.

Due to the condition $E\ll m \ll U \ll \Upsilon$, we may neglect $E$ and $m$ inside the nucleus and estimate classical momentum using Eqs. (\ref{eqphi},\ref{V}): $p^2 \approx U \Upsilon $. Assuming constant nuclear charge  density $\rho= 3 Z |e|/ (4\pi R^3)$, we obtain 
\begin{equation} \label{UpsilonU}
\Upsilon=-\frac{3 Z \alpha}{m^2  R^3}\,\, , U= -\frac{Z \alpha}{R}(\frac{3}{2} - \frac{r^2}{2 R^2}) \,. 
\end{equation}
Integration in Eq. (\ref{BohrIn}) gives the principal quantum number shift 
\begin{equation} \label{Delta}
\Delta \approx  0.6 \frac{ Z \alpha}{mR} \gg 1.
\end{equation}
This means that there are many orbitals confined to the nucleus, on the distance $r \lesssim R$. For $R\to 0$ the number of such states tends to infinity. This result agrees with the conclusion of Ref. \cite{Kuchiev2008}. 

\subsection{Vector particle states inside the nucleus}
 Condition $mR \ll  Z \alpha \ll 1$ gives us $m \ll U \ll \Upsilon$ inside the nucleus. However, the uncertainty relation for the localised states, $p \gtrsim 1/R$, indicates that we can not neglect $\epsilon$. 
 Then the substitution of the semiclassical wave function (\ref{vclassical}) to Eqs. (\ref{eqphi},\ref{V}) gives   
$p^2 \approx \Upsilon(U- \epsilon)$, and  the Bohr-Sommerfeld  quantization equation
\begin{equation} \label{BohrInside}
\int_0^{r_t} [\Upsilon(U -\epsilon)]^{1/2} dr \approx \pi n,
\end{equation}
This equation  is valid if the classical turning point $r_t <R$, equivalent to $\epsilon < - Z \alpha/R$. Calculation of this integral gives 
\begin{equation} \label{EpsilonInside}
\epsilon \approx  - \frac{3 Z \alpha}{2R} + \sqrt{\frac{8}{3}} m \,n \,,
\end{equation}
The condition $\epsilon < - Z \alpha/R$ gives 
\begin{equation} \label{nInside}
n < 0.3 \frac{Z \alpha}{mR} \,\,.
\end{equation}
Levels with higher $n$ may have a tail outside the nucleus.  For $R\to 0$ the number of levels inside the nucleus tends to infinity.

At large distance , when potential $U$ may be neglected, wave equation depends on $\epsilon^2$, so both $\epsilon >  m$ and $\epsilon <-m$ solutions belong to the continuum spectrum. 
This means that all states described by  Eq. (\ref{EpsilonInside}) are not true  eigenstates with real values of  energy but they are resonances with a large width corresponding to complex energy  poles in the $W^+$ scattering amplitude
( compare with  similar  complex energy electron states with  $\epsilon < - m$ in nuclei with charge $Z>170$ described in the review \cite{ZeldovichPopov}).  In the formal solution of the wave equations, wave functions of such resonance states at real energies Eq. (\ref{EpsilonInside}) have oscillating long distance tail instead of exponential decay. Near the resonances positions described by Eq. (\ref{EpsilonInside}) the ratio of the squared wave function at  small distance divided by the squared wave function amplitude  at large distance has the Breit-Wigner type enhancement. Note that for states with  $\epsilon < - m$, the potential  $U$ presents a potential barrier since it appears in the wave equations as a combination $\epsilon -U$.

  Existence of states with  $Re(\epsilon) < - m$  corresponds to  creation of particle-antiparticle pairs, the vacuum breakdown. In a dynamical picture, negatively charged particles, produced this way,  neutralise  nuclear charge, stopping further pair production.  In other words, the nuclear charge is screened completely by the particles from vacuum.

Our estimates are based on the inequality $\Upsilon \gg U$. However, the accuracy of our semiclassical estimate of the energy levels may be affected by the singularity  $\sim 1/(\epsilon-U)$ in the wave equations (\ref{eqphi},\ref{V},\ref{GH}), which is close to the classical turning point $\epsilon-m=U$. However, this singular term only slightly shifts the classical turning point $p=0$. The interval near the turning point, where neglected terms are significant, is small, $x \sim (mR/Z \alpha)^{2/3} R$, and excluding this interval  does not affect the Bohr-Sommerfeld integral significantly, if $n \gg 1$.  Similar argument justifies neglecting the centrifugal term $\sim 1/r^2$.

For further analysis, it is convenient to present the wave equations in the two-component form ($v,w$), which looks similar to the radial Dirac equations for electron.
\begin{equation}\label{vw}
\begin{aligned}
&\left[(\epsilon-U)^2-m^2\right]v=-(\epsilon -U)w' -U' w \,,\\
&(\epsilon -U -\Upsilon)w=\frac{2v}{r} + v' \,.
\end{aligned}
\end{equation}
From these equations, we see that $v$ and $w$ have no singularities at $(\epsilon -U)=0$, they pass this point smoothly. Similar to the Dirac equation, such singularity appears  only when  we substitute  the lower component $w$ into the equation for the upper component $v$ and divide equation by $(\epsilon -U)$, to have coefficient before $v''$ equal to 1. 

Note that the lower component $w$ in the non-relativistic limit $\epsilon \approx m \gg U$ at large distance $r$ is proportional to $(p/m)v$, similar to the lower component in the Dirac equation.

\subsection{Vacuum polarization}\label{Vacuum polarization}

In the case of point-like nucleus, the vacuum polarization potential at small distance $r m_z\ll 1$ ($m_z$ is the maximal mass of the vacuum particle)  may be presented as 
\begin{equation} \label{vacuum}
U_v(r)= [\alpha \beta \ln(m_z r)] \frac{Z \alpha}{r}\,, 
\end{equation}
where $\beta$ is the lowest order coefficient of the  Gell-Mann - Low $\beta$-function.
This gives for $r>0$
\begin{equation}\label{Upsilonv}
\Upsilon_v=\frac{e\rho_v}{m^2}=-\frac{\Delta U_v}{m^2} = \frac{Z\alpha^2 \beta}{m^2 r^3} \,.
\end{equation}
In QED $\beta>0$ and $\Upsilon$ produces impenetrable  barrier which prevents collapse of the vector particle wave function \cite{Kuchiev2006}.  However, in asymptotically free theories with $\beta<0$ there is no such barrier. 

In addition to the polarization charge  distribution outside the nucleus, there is a $\delta(r)$ term (from $\Delta (1/r)$ factor in Eqs. (\ref{vacuum},\ref{Upsilonv})) bringing the total vacuum polarization charge to zero. Integral of the polarization charge outside the nucleus is  logarithmically divergent as $\int dr/r$, so the coefficient before the $\delta(r)$ is infinite. One can also see this from the divergent at $r \to 0$ factor  $\ln(m_z r)$ in  Eqs. (\ref{vacuum},\ref{Upsilonv}). 

Finite nuclear radius $R$ eliminates this divergency and produces $\Upsilon_v$ inside the nucleus which has opposite sign to that of $\Upsilon_v$ outside the nucleus. In the QED case with $\beta>0$ the particle penetrates the $\Upsilon_v$ barrier and falls to the deep potential well inside the nucleus, while in the asymptotically free case $\beta<0$ the vacuum polarization barrier appears inside the nucleus.

\subsection{Charge density}\label{Charge density}

Using Eqs.~\eqref{RhoDef} and \eqref{LongDef}, we write the charge density as
\begin{equation}\label{LongRhoDef}
\rho^W=2e\left[(\epsilon-U)(v^2+w^2)+2v\frac{dw}{dr}-\Upsilon w^2\right]\,,
\end{equation}
where $w$ is defined as
\begin{equation}\label{wDef}
w\equiv\frac{1}{\epsilon-U  -\Upsilon}\left(\frac{dv}{dr}+\frac{2v}{r}\right)\,.
\end{equation}

For a point-like nucleus, in the relativistic region $r\ll \zeta/m$, the wave function $v$ behaves as $v\sim r^{L-2}$  so
\begin{equation}
w\sim Lr^{L-2}\,,
\end{equation}
and, as a result, in this regime, the charge density \eqref{LongRhoDef} behaves as (recall that $L=\sqrt{1/4-\zeta^2}+1/2$)
\begin{equation}
\rho^W\sim-2eL(3-2L)r^{2L-5}/\zeta\,.
\end{equation}
This charge density diverges so badly as $r\rightarrow 0$ that the total chare $Q=\int\rho^W d^3r$ localized inside any small sphere around the nucleus is infinite. It is also of a "wrong" sign (positive for a negatively charged vector particle), as pointed out in Ref.~\cite{Flambaum2007}. This looks like the integral of the charge density has opposite sign to the particle charge $e$.  

 Let us now consider our case of the finite nucleus. As we have shown above, there are states localised mainly inside the nucleus. However, we will show that even for the states having Schrodinger-like spectrum, a finite fraction of  their  charge is located inside the nucleus  for $R\to 0$.
 
 Let us firstly estimate the contribution of the charge density without $\Upsilon$ term. 
 A singular part of the vector particle charge density outside may be presented  as $\rho^W(r) = \rho^W(R)(r/R)^{2 \gamma -4}$. Charge density inside is $\rho^{no \Upsilon}(r) \approx \rho^W(R)$. 
  Integrating charge density inside from 0 to $R$ and charge density outside from $R$  to infinity with $r^2 dr$ we obtain   
\begin{equation}
\frac{Q^{no \Upsilon}_{\rm in}}{Q_{\rm out}} \approx  \frac{1 - 2 \gamma}{3} \approx \frac{2 Z^2 \alpha^2}{3}.
\end{equation}
We see that this ratio does not depend on $R$, i.e. vector particle has a finite probability to be inside the nucleus  even in the case when the nucleus has zero size. Also, the both charges inside and near the nucleus have wrong sign.


Now include the $\Upsilon=-\frac{3 Z \alpha}{m^2 R^3} \gg U$  contribution to $w$. This gives the ratio of the charge inside to the charge outside as 
\begin{equation}
\frac{Q_{\rm in}}{Q_{\rm out}} \sim  -\frac{  2 (Z \alpha)^4}{3}\,,
\end{equation}
which stays  finite   for $R \to 0$.
This means that even for the states with the  Schrodinger-type spectrum  their charge has finite fraction inside the nucleus, if $mR \ll Z \alpha$. 

Note while the particle charge density changes sign in a vicinity of the nucleus and $\rho_W$ becomes positive,  $\rho_W$ stays negative  inside the nucleus due to the large  $\Upsilon$ term in the denominator of Eq.  (\ref{wDef}) suppressing  $w$ inside the nucleus.


\section{Summary}\label{Conclusion}

 The Coulomb bound–state problem for a charged \emph{vector} particle is
pathological at short distances: its wave function collapses onto the
Coulomb centre  already for infinitesimal coupling
($\zeta = Z\alpha \to 0$))~\cite{Schwinger1940}.  By contrast, the critical charge for the point-like nucleus 
is \(Z\alpha=1/2\) for a spin-0 particle and \(Z\alpha=1\) for a
spin-\(1/2\) particle.  

 To investigate this behavior further,  we have solved the problem for a light charged vector particle in the field of a finite-size, infinitely heavy nucleus, neglecting electroweak
corrections and nuclear recoil.  The point-like limit is recovered by
taking $R\to 0$. This approach allows us to include a critically important $\Upsilon$ term in the wave equations, which for the point-like nucleus is proportional to $\delta(r)$ and was ignored in the previous calculations of the energy spectrum. 

We focus on the most interesting case of the longitudinal polarization with the  angular momentum $j=0$, and a small nuclear radius,  $mR \ll Z\alpha \ll 1$. Note that the Coulomb  potential inside the nucleus in this case exceeds mass, $Z\alpha/R > m$. This is a typical situation for atomic electrons and does not create any  problem, the Schrodinger  equation still gives correct results. However, for spin 1 particles the difference  with the Schrodinger  equation solution is profound, even  for a very  small interaction constant $Z \alpha$ 
(there is agreement with the Schrodinger equation results in the true non-relativistic case $Z\alpha/R \ll  m$, when $\Upsilon \ll U \ll m$).

 The spectrum splits naturally into two parts. A set of “deep” states lies predominantly inside the nucleus. Their number grows as $N \sim 0.6 Z \alpha/mR$, so $N \to \infty$ as the nuclear radius $R$ shrinks to zero. The energies of these intranuclear states (in Gaussian units) are
 \begin{equation} \label{EpsilonInsideC}
\epsilon \approx  - \frac{3 Z e^2}{2R} + \sqrt{\frac{8}{3}} m c^2  \,n \,,
\end{equation}
 which places them deep in the negative‑energy continuum, $\epsilon  \sim - Z e^2/ R < - m c^2$ \footnote{Wave functions of  states with $\epsilon   < - m c^2$ do not vanish at large distance. These states are not true  bound states with real values of  energy but they are resonances with a finite  width corresponding to complex energy  poles in the $W^+$ scattering amplitude.}
. Physically, this means the Coulomb field becomes strong enough to pull particle–antiparticle pairs out of the vacuum (“vacuum breakdown”).
The negative charges produced in this way screen the nuclear charge, halting further pair creation and driving the system toward complete charge neutralisation
 \footnote{The negative energy states and vacuum breakdown may be eliminated by the additional Compton-like diagram, which appears in a renormalisable model, in which  the heavy particle producing Coulomb potential has a neutral partner.  This diagram produces  repulsive potential on the distance $r \lesssim 1/M$, where $M$ is the mass of the partner ~\cite{flambaum2006}. Nevertheless, the wavefunction still exhibits pathological behaviour, collapsing to $r \sim 1/M$, where $M$ may be arbitrary large.}.

The second part of the spectrum (in Gaussian units) is given by
\begin{equation} \label{CoulombEnergyC}
E_n=- \frac{m Z^2 e^4}{2 \nu^2},
\end{equation}
where $E_n =\epsilon - mc^2$ is the non-relativistic energy, $ \nu=n-\Delta$ is the effective principal quantum number, and 
\begin{equation} \label{DeltaC}
\Delta \approx  0.6 \frac{ Z \alpha}{mR} \gg 1.
\end{equation}
 The magnitude of $\Delta =n- \nu $ indicates the number of levels  with $j=0$ inside the nucleus. Parameter $\Delta$ and the effective principal quantum number $\nu=n-\Delta$  are  not integer numbers, so the energy levels do not coincide with the (re-numbered) energy levels in the Coulomb potential with a point-like nucleus. This formula is similar to the Rydberg formula for  energy of  external electron traveling far from atomic core in the Coulomb field of the remaining ion.

 The ratio of the charge inside the nucleus to the charge outside the nucleus for such Sommerfeld-type levels  is estimated as 
 \begin{equation}\label{Q} 
\frac{Q_{\rm in}}{Q_{\rm out}} \sim  -\frac{  2 (Z \alpha)^4}{3}\,,
\end{equation}
which stays  finite   for $R \to 0$.
This means that  the  charge of the states with "ordinary" Sommerfeld-type spectrum  has finite fraction inside the nucleus, even for infinitely small nucleus.

 Ref.~\cite{Flambaum2007} found that  the charge density $\rho_W$ for a negatively charged vector particle near the nucleus changes sign and becomes positive. This effect is due to the dominating contribution of the vector boson's electric quadrupole moment and spin-orbit interaction in the area of rapidly varying wave function. Since the integral of this wrong sign density is divergent for  a point-like nucleus,  this looks like a paradox, charge of a vector particle in the Coulomb field changes sign.
 However, we found that for finite $R$, the  charge density  $\rho_W$  is negative inside the nucleus (due to the $\Upsilon$ term contribution).

Vacuum polarization modifies the Coulomb field at $r \ll 1/m_z$ as $U_v(r)= [\alpha \beta \ln(m_z r)] Z \alpha/r$, inducing $\Upsilon_v(r)= Z\alpha^2 \beta/m^2 r^3$ plus a compensating $\delta(r)$ term so the total induced charge vanishes. For QED ($\beta>0$) this gives a repulsive $1/r^3$ barrier outside the nucleus that suppresses short-distance collapse; a finite nuclear radius $R$ regularizes the divergence and produces an opposite-sign $\Upsilon_v$ inside, effectively creating a deep inner well accessible by penetration. In asymptotically free theories ($\beta<0$) the exterior barrier is absent and a barrier appears inside the nucleus.

These results show that the Coulomb problem for spin-1 particles
requires special regularisation:  the probability density exhibits an unavoidable short-distance collapse
unless stabilised by the finite nuclear mass, finite nuclear size or by additional
short-range interactions. Note that the Higgs boson  exchange produces an attractive potential which can not provide a cure.

Further investigation has shown that there is no collapse of the $W$ wave function in either the Lorentz-scalar Yukawa potential or in the gravitational field
\footnote{Non-stationary quasibound states of vector particles in the Schwarzschild black hole metric were found in Ref. \cite{Rosa}. To consider stationary states, we studied  a near–black-hole metric in which the radius of the central body $R$ slightly exceeds the gravitational radius $r_g$, and then took the limit $R \to r_g$. This approach parallels that of Refs.~\cite{BlackHole1,BlackHole2,BlackHole3,BlackHole4}, where the cases of massless vector particles (photons) as well as massive fermions and scalars were examined. These works reported a dense spectrum of resonances and  - for massive particles - bound states inside the body (with the spectrum becoming infinitely dense as $R \to r_g$). However, this spectrum is unrelated to the collapse of the wave function observed for vector particles in the Coulomb field. 
In particular, we find no qualitative difference between  a scalar field and the $j=0$ component of a vector field in this respect.}. 
We therefore conclude that wave-function collapse is specific to vector-mediated interactions (Coulomb photon exchange and neutral weak $Z$ exchange), and does not occur for Lorentz–scalar or gravitational interactions.

\vspace{2mm}
\textit{Acknowledgements.}--- 
We are  grateful to Raghda  Abdel Khaleq for her contribution during the initial stage of this work.
 The work was supported by the Australian Research Council Grant No.\ DP230101058.
 \appendix

\section{Magnetic polarization, \texorpdfstring{$j\geq 1$}{}}
In this section, we consider the magnetically polarized mode. Substituting Eq.~\eqref{MagDef} into Eq.~\eqref{FinalLagrange}, we obtain the following equation for the radial wave function $f$  
\begin{equation}\label{MagEq}
 \left[\Delta_j+(\epsilon-U)^2-m^2\right]f=0\,,
\end{equation}
where 
\begin{equation}
    \Delta_j\equiv\frac{1}{r^2}\frac{d}{dr}\left(r^2\frac{d}{dr}\right)-\frac{j(j+1)}{r^2}.
\end{equation}
One recognizes Eq.~\eqref{MagEq} as the Klein-Gordon equation. 

For simplicity, we assume a model in which $U$ is a Coulomb potential with a sharp cut off at the nuclear radius $R$
\begin{equation}\label{PotUSharp}
U=-\frac{\zeta}{R}\theta(R-r)-\frac{\zeta}{r}\theta(r-R)\,,
\end{equation}
where $\zeta\equiv Z\alpha$. This potential, which is produced by a uniform distribution of the nuclear charge on the surface of the nucleus, allows for an analytical solution.

Inside the nucleus, where the potential is constant, $U=-\zeta/R$, the solution to Eq.~\eqref{MagEq} which is regular at $r=0$ reads
\begin{equation}\label{MagSolIn}
f_{\rm in}=j_j(kr)\,,
\end{equation}
where $k\equiv\sqrt{(\epsilon+\zeta/R)^2-m^2}$ and $j_j$ is the spherical Bessel wave function of the first kind.

Outside the nucleus, where $U=-\zeta/r$, a general solution to Eq.~\eqref{MagEq} has the form
\begin{equation}\label{MagWFunc}
\begin{aligned}
    f=\left[c_1M_{\lambda,\gamma}(2\beta r)+c_2W_{\lambda,\gamma}(2\beta r)\right]/r\,,
\end{aligned}
\end{equation}
where $\beta\equiv \sqrt{m^2-\epsilon^2}$, $\lambda\equiv \zeta\epsilon/\sqrt{m^2-\epsilon^2}$, $\gamma\equiv\sqrt{(j+1/2)^2-\zeta^2}$. Here, $M_{\lambda,\gamma}$ and $W_{\lambda,\gamma}$ are Whittaker functions of the first and second kind, respectively.

In the pointlike case, since $W_{\lambda,\gamma}(2\beta r)$ is singular at $r=0$, it must be excluded from the solution \eqref{MagWFunc}. Requiring the function $M_{\lambda,\gamma}(2\beta r)$ to be finite at $r=\infty$ then results in the Sommerfeld formula for the energy eigenvalue
\begin{equation}\label{MagSommerfeld}
\epsilon_{nj}=m\left[1+\frac{\zeta^2}{\left(\gamma+n-j-1/2\right)^2}\right]^{-1/2}\,,
\end{equation}
where $n=1,2,...$ is the principle quantum number.

In the case of a non-pointlike nucleus, the solution \eqref{MagWFunc} is valid only for the region outside the nucleus, where $r\neq 0$, so $W_{\lambda,\gamma}(2\beta r)$ is admissible and automatically satisfies the finiteness condition at infinity. On the other hand, since the energy $\epsilon$ receives corrections due to the finite size of the nucleus, it no longer satisfies the Sommerfeld condition \eqref{MagSommerfeld}. As a result, $M_{\lambda,\gamma}(2\beta r)$ is now irregular at $r=\infty$ and must thus be excluded from Eq.~\eqref{MagWFunc}, giving 
\begin{equation}\label{MagSolOut}
f_{\rm out}=W_{\lambda,\gamma}(2\beta r)/r\,.
\end{equation}


By matching the logarithmic derivatives of the two solutions \eqref{MagSolIn} and \eqref{MagSolOut} at the nuclear boundary $r=R$, one is able to find the nuclear size correction to the Sommerfeld energy \eqref{MagSommerfeld}. Since we assume that $mR,\zeta\ll1$, the parameter $kR$ is always small\footnote{For later use, we note in passing that in the relativistic limit $\zeta/R\gg m$, $kR\approx\zeta$ whereas in the nonrelativistic limit $\zeta/R\ll m$, $kR\approx\sqrt{2\zeta mR}$.} and we may find the logarithmic derivative of the ``inside'' wave function as
\begin{equation}\label{MagLogDerivIn}
\left.\frac{f'_{\rm in}}{f_{\rm in}}\right|_{r=R}=\frac{\xi_j(R)}{R}\,,
\end{equation}
where
\begin{equation}\label{xij}
\begin{aligned}
\xi_j(R)&\equiv\frac{kR\left[j_{j-1}(kR)-j_{j+1}(kR)\right]}{2j_j(kR)}-\frac{1}{2}\\
&\approx j-\frac{(kR)^2}{2j+3}\,.
\end{aligned}
\end{equation}


On the other hand, outside the nucleus, since $2\beta R =2\sqrt{m^2-\epsilon^2}R<2mR\ll 1$, for $r\approx R$, we can make use of the following expansion of the Whittaker function
\begin{equation}\label{ExpandWhittaker}
\begin{aligned}
W_{\lambda,\gamma}(x)&\approx \frac{\Gamma(-2\gamma)x^{\gamma+\frac12}}{\Gamma(-\gamma+1/2-\lambda)}\left(1-\frac{\lambda x}{2\gamma+1}\right)\\
&+\frac{\Gamma(2\gamma)x^{-\gamma+\frac12}}{\Gamma(\gamma+1/2-\lambda)}\left(1+\frac{\lambda x}{2\gamma-1}\right)\,,
\end{aligned}
\end{equation}
and write
\begin{equation}\label{MagExpandWhittaker}
\begin{aligned}
f_{\rm out}&\approx r^{\gamma-\frac12}\left(1-\frac{2\zeta\epsilon r}{2\gamma+1}\right)\\
&+\frac{\Gamma(2\gamma)\Gamma(-\gamma+1/2-\lambda)}{\Gamma(-2\gamma)\Gamma(\gamma+1/2-\lambda)}\frac{r^{-\gamma-\frac12}}{(2\beta)^{2\gamma}}\left(1+\frac{2\zeta\epsilon r}{2\gamma-1}\right)\,,
\end{aligned}
\end{equation}
where we have taken out the common factor $\Gamma(-2\gamma)(2\beta)^{\gamma}/\Gamma(-\gamma+1/2-\lambda)$. 

At a first glance, it appears that by setting $r=R$ and taking the limit $R\rightarrow 0$, the second term in Eq.~\eqref{MagExpandWhittaker} is singular. However, we point out that as $R\rightarrow 0$, the energy $\epsilon$ assumes the Sommerfeld form \eqref{MagSommerfeld}, which is equivalent to $\gamma+1/2-\lambda=-n+j+1=0,-1,...$ so the function $\Gamma(\gamma+1/2-\lambda)=\Gamma(-n+j+1)$ becomes infinite. This singularity of the gamma function necessarily cancels that of $R^{-\gamma-1/2}$, making the wave function $f$ finite. In fact, as $R\rightarrow 0$, the solution $W_{\lambda,\gamma}(\rho)$ becomes $M_{\lambda,\gamma}(\rho)$.

In order to account for this behavior of the wave function $f_{\rm out}$, we expand the gamma function around its pole as
\begin{equation}\label{ExpandGammaFunc}
\begin{aligned}
\Gamma(\gamma+1/2-\lambda)&\approx\Gamma\left(\gamma+1/2-\lambda(\epsilon_{nj})-\frac{d\lambda}{d\epsilon}\Delta\epsilon\right)\\
&=\Gamma\left(-n+j+1-\frac{\zeta m^2\Delta\epsilon}{(m^2-\epsilon^2_{nj})^{\frac32}}\right)\\
&\approx\frac{(-1)^{n-j}}{(n-j-1)!}\frac{(m^2-\epsilon^2_{nj})^{\frac32}}{\zeta m^2\Delta\epsilon}\,,
\end{aligned}
\end{equation}
where we have made use of the approximation
\begin{equation}
\Gamma(-n+x)\approx\frac{(-1)^n}{n!x}\,.
\end{equation}
Substituting Eq.~\eqref{ExpandGammaFunc} into Eq.~\eqref{MagExpandWhittaker} and setting $\epsilon\approx\epsilon_{nj}$ everywhere, one obtains
\begin{equation}\label{ApproxfOut}
\begin{aligned}
f_{\rm out}&\approx r^{\gamma-\frac12}\left(1-\frac{2\zeta\epsilon_{nj}r}{2\gamma+1}\right)\\
&+\frac{\Gamma(-2\gamma-n+j+1)}{\Gamma(-2\gamma)}\frac{(n-j-1)!\Gamma(2\gamma)}{(-1)^{n-j}}\\
&\times\frac{\zeta m^2\Delta\epsilon}{\beta_{nj}^3}\frac{r^{-\gamma-\frac12}}{(2\beta_{nj})^{2\gamma}}\left(1+\frac{2\zeta\epsilon_{nj}r}{2\gamma-1}\right)\,,
\end{aligned}
\end{equation}
where $\beta_{nj}\equiv\beta(\epsilon_{nj})=\sqrt{m^2-\epsilon_{nj}^2}$. It is clear from Eq.~\eqref{ApproxfOut} that for $f_{\rm out}$ to remain finite as $R\rightarrow0$, $\Delta\epsilon$ must vanish at least as fast as $R^{2\gamma}$. Using Eq.~\eqref{ApproxfOut}, one finds
\begin{widetext}
\begin{equation}\label{MagLogDerivOut}
\left.\frac{f'_{\rm out}}{f_{\rm out}}\right|_{r=R}\approx\frac{\gamma-\frac12-\zeta\epsilon_{nj}R-\frac{\Gamma(-2\gamma-n+j+1)}{\Gamma(-2\gamma)}\frac{(n-j-1)!\Gamma(2\gamma)}{(-1)^{n-j}}\frac{\zeta m^2\Delta\epsilon}{\beta_{nj}^3}\frac{R^{-2\gamma}}{(2\beta_{nj})^{2\gamma}}\left(\gamma+\frac12+\zeta\epsilon_{nj}R\right)}{R\left[1-\frac{2\zeta\epsilon_{nj}R}{2\gamma+1}+\frac{\Gamma(-2\gamma-n+j+1)}{\Gamma(-2\gamma)}\frac{(n-j-1)!\Gamma(2\gamma)}{(-1)^{n-j}}\frac{\zeta m^2\Delta\epsilon}{\beta_{nj}^3}\frac{R^{-2\gamma}}{(2\beta_{nj})^{2\gamma}}\left(1+\frac{2\zeta\epsilon_{nj}R}{2\gamma-1}\right)\right]}\,.
\end{equation}
\end{widetext}

Equating the logarithmic derivatives of the ``inside'' and ``outside'' solutions, Eqs.~\eqref{MagLogDerivIn} and \eqref{MagLogDerivOut}, one may solve for the energy correction $\Delta\epsilon$ as
\begin{equation}\label{MagDeltae}
\begin{aligned}
\Delta\epsilon&\approx\frac{(-1)^{n-j}\Gamma(-2\gamma)(2\beta_{nj}R)^{2\gamma}}{(n-j-1)!\Gamma(2\gamma)\Gamma(-2\gamma-n+j+1)}\\
&\times\left[\frac{\gamma-\xi_j-\frac12}{\gamma+\xi_j+\frac12}+\frac{(2\xi_j-2\gamma-1)\zeta\epsilon_{nj}R}{(2\gamma+1)(\gamma+\xi_j+\frac12)}\right]\frac{\beta_{nj}^3}{\zeta m^2}\,.
\end{aligned}
\end{equation}
We observe that the leading term in Eq.~\eqref{MagDeltae} scales as $R^{2\gamma}$, in agreement with the requirement that the ``outside'' wave function is regular at $r=R\rightarrow0$.

We point out, however, that this leading behavior of the energy correction is strictly relativistic. Indeed,  it may be verified that in the nonrelativistic limit where $\zeta\ll mR\ll 1$, the term in Eq.~\eqref{MagDeltae}, which is proportional to $R^{2\gamma}$, vanishes and the remaining term reduces to 
\begin{equation}\label{MagNonRel}
\Delta\epsilon\approx\frac{(n+l)!}{n^{2l+1}(n-l-1)!}\frac{(2\zeta mR)^{2l+2}}{(2l+3)!(2l+1)!}\frac{\zeta^2 m}{n^3}\,,
\end{equation}
where $l=1,2,...$ is the vector boson's orbital angular momentum (for the magnetic mode, $j=l$). Equation~\eqref{MagNonRel} agrees with the nonrelativistic result~\eqref{PerbTheoResult} obtained by applying first-order perturbation theory to the Schr\"odinger equation

\section{Energy correction by perturbation theory}

Assuming the model of a hollow nuclear shell, Eq.~\eqref{PotUSharp}, the perturbation to the Coulomb potential in the Schr\"odinger equation may be written as
\begin{equation}\label{Perturbation}
V=\left(\frac{\zeta}{R}-\frac{\zeta}{r}\right)\theta(R-r)\,,
\end{equation}
where $\theta$ is the Heaviside step function.

Since the perturbation \eqref{Perturbation} is localized inside the nucleus, in the limit where the nuclear radius $R$ is much smaller than the Bohr radius $a_B$, the unperturbed wave function corresponding to the principal quantum number $n$ and angular momentum $l$ may be written as
\begin{equation}\label{UnperturbedWave}
\varphi\underset{\begin{smallmatrix} 
 r\ll a_B 
\end{smallmatrix}}{\mathop{\approx }}\sqrt{\left(\frac{2Z}{na_B}\right)^3\frac{(n+l)!}{2n(n-l-1)!}}\frac{1}{(2l+1)!}\left(\frac{2Zr}{na_B}\right)^l\,.
\end{equation}

Using Eqs.~\eqref{Perturbation} and \eqref{UnperturbedWave}, we find the energy correction as 
\begin{equation}\label{PerbTheoResult}
\begin{aligned}
\Delta\epsilon&=\bra{\varphi}V\ket{\varphi}\approx\left(\frac{2Z}{na_B}\right)^3\frac{(n+l)!}{2n(n-l-1)!}\\
&\times\frac{1}{[(2l+1)!]^2}\int_0^R\left(\frac{2Zr}{na_B}\right)^{2l}\left(\frac{\zeta}{r}-\frac{\zeta}{R}\right)r^2dr\\
&=\frac{(n+l)!}{n^{2l+1}(n-l-1)!}\frac{(2\zeta mR)^{2l+2}}{(2l+3)!(2l+1)!}\frac{\zeta^2 m}{n^3}\,.
\end{aligned}
\end{equation}

\bibliography{references}
\end{document}